# Photonic flywheel in a monolithic fiber resonator

Kunpeng Jia[1,2]†, Xiaohan Wang[1,2]†, Dohyeon Kwon[3], Jiarong Wang[2], Eugene Tsao[2], Huaying Liu[1,2], Xin Ni[1], Jian Guo[1], Mufan Yang[1], Xiaoshun Jiang[1], Jungwon Kim[3], Shi-ning Zhu[1], Zhenda Xie[1]*, Shu-Wei Huang[2]*

[1] National Laboratory of Solid State Microstructures, School of Electronic Science and Engineering, College of Engineering and Applied Sciences and School of Physics, Nanjing University, Nanjing 210093, China.

[2] Department of Electrical, Computer and Energy Engineering, University of Colorado Boulder, Boulder, CO 80309, United States.

[3] Department of Mechanical Engineering, Korea Advanced Institute of Science and Technology, Daejeon 34141, Republic of Korea.

We demonstrate the first compact photonic flywheel with sub-fs time jitter (averaging times up to 10 μs) at the quantum-noise limit of a monolithic fiber resonator. Such quantum-limited performance is accessed through novel two-step pumping scheme for dissipative Kerr soliton (DKS) generation. Controllable interaction between stimulated Brillouin lasing and Kerr nonlinearity enhances the DKS coherence and mitigate the thermal instability challenge, achieving a remarkable 22-Hz intrinsic comb linewidth and an unprecedented phase noise of -180 dBc/Hz at 945 MHz carrier at free running. The scheme can be generalized to various device platforms for field-deployable precision metrology.

Dissipative Kerr soliton (DKS) frequency comb [1, 2], generated by pumping a high-Q monolithic resonator with a resonant continuous-wave (cw) single-mode laser, has recently emerged as a promising complement to the traditional mode-locked laser frequency comb [3-6]. Access to considerably larger comb spacings in non-conventional spectral ranges has led to continued enthusiasm since its advent. Demonstrations of photonic frequency synthesizer [7, 8], microphotonic astrocomb [9, 10], dual comb spectroscopy [11, 12], and coherent optical communication [13, 14] revealed the unique performance of DKS frequency comb, and reassured further expansion of already remarkable applications. However, despite the massive achievements in the spectral domain, noise analysis of DKS in the time domain is still lacking and thus it remains questionable whether DKS can also serve as an optical flywheel [5], where the pristine temporal periodicity with sub-optical-cycle timing jitters can be utilized for intriguing applications at the intersection of ultrafast optics and microwave electronics, including photonic analog-to-digital converters (ADCs) for next-generation radar and communication systems [15-17], coherent waveform synthesizers for pushing the frontiers of femtosecond and attosecond science [18-20], ultra-low noise microwave signals generation [21-25], and timing distribution links for synchronizing large-scale scientific facilities like X-ray free electron lasers and the extreme light infrastructure [26-32].

On-chip monolithic resonators feature strong mode confinement and large Kerr nonlinearity that fundamentally elevate the quantum-limited phase noise and prohibit us from achieving timing jitter at the optical flywheel level [2, 33]. Moreover, sophisticated pump noise control and thermal effect management are required to practically achieve the quantum-limited phase noise. As pump is an integral part of the DKS frequency comb [1], narrow-linewidth pump is necessary for reducing the pump-to-comb noise conversion to reach the quantum-limited comb coherence [34]. Mitigation of the strong thermal nonlinearity is also a fundamental challenge for stable DKS generation. Traditionally, either active precision control of pump and resonators is implemented to reduce the thermal effect [35, 36] or an auxiliary laser is included to compensate for the pump induced thermal dynamics [37, 38]. Both mitigation schemes inevitably add system complexity and compromise the advantages of the DKS frequency comb as an integrated device.

In this work, we demonstrate for the first time a compact photonic flywheel based on DKS generation in a relatively unexplored platform of monolithic fiber Fabry-Perot (FFP) resonator [39-41]. Our theoretical noise analysis suggests that FFP resonator is a promising photonic flywheel platform where sub-femtosecond and even 100-attosecond timing jitters can be achieved at gigahertz repetition rate. To mitigate challenges in DKS generation and reach the quantum-limited performance, we devise a novel two-step pumping scheme where the DKS is formed from a secondary pump generated in the same FFP resonator through the cross-polarized stimulated Brillouin lasing (SBL). Advantages of such two-step pumping scheme are two-fold. First, resonator SBL exhibits an orders-of-magnitude linewidth reduction [42, 43] and thus as the secondary pump, it has a much higher spectral purity in comparison to the primary laser, reducing the pump-to-comb noise conversion [34] and relaxing the laser linewidth requirement to achieve a DKS with quantum-limited phase noise and timing jitter. Second, the primary laser and the SBL work together to compensate the detrimental cavity thermal nonlinearity that limits the reliability and robustness of DKS generation [2]. Only one cw single-mode laser is required in our two-step pumping scheme, where the primary laser evolves to effectively become an auxiliary laser while the cross-polarized SBL grows to generate DKS in the FFP resonator, resulting in a novel self-stabilized strategy for thermal management and robust DKS generation. All in all, our free-running quantum-limited DKS frequency comb achieves an unprecedented intrinsic linewidth of 22 Hz, characterized by a short delay self-heterodyne interferometry (SDSHI) [44], and breaks for the first time into the attosecond integrated timing jitter realm of 995 attosecond for averaging times up to 10 μs, measured with an all-fiber reference-free Michelson interferometer (ARMI) timing jitter measurement apparatus [45].

Our FFP resonator (inset of Fig. 1(a)) is fabricated from a commercially available highly nonlinear fiber (HNLF, NL-1550-POS, YOFC) with nonlinear coefficient over 10 $W^{-1}km^{-1}$ and attenuation below 1.5 dB/km. The 105-mm long cavity length results in a free spectral range (FSR) of 945.4 MHz as shown in Fig. 1(b). Both fiber ends are mounted in ceramic ferrules, finely polished and then coated with 10-layer pairs of $Ta_2O_5$ and $SiO_2$ using ion assisted deposition method. Such Bragg mirror achieves a reflectivity of over 99.6 % from 1530 nm to 1570 nm. Of note, gigahertz comb spacing is especially suitable for applications like dual-comb spectroscopy and photonic ADCs [46, 47]. With negligible propagation loss of HNLF, the quality factor (Q) of FFP resonator is mainly defined by the dielectric mirror coatings at the two ends. Thus its Q and the comb generating pump power both scale favorably at lower comb spacings, which is critically different from on-chip monolithic resonators with propagation-limited Q. The FFP resonator resonance linewidth is measured and fitted to be 5.6 MHz, corresponding to a Q of $3.4\times10^7$ (inset of Fig. 1(b)). Group velocity dispersion (GVD) of -3 $fs^2$/mm at 1557 nm is characterized with the spectral interferometric method [48]. Similar monolithic FFP resonators have been previously utilized for Brillouin-enhanced hyperparametric frequency comb generation [39] and synchronously pulse pumped DKS generation [40]. However, our work presents the first demonstration of cw pumped DKS generation in a monolithic FFP resonator, which is the key to the photonic flywheel demonstration.

In an FFP resonator, there are two families of cavity modes with orthogonal polarizations (P1 and P2) due to the stress-induced birefringence (Fig. 1(b)) [49]. The offset frequency $\Delta f_{P1-P2}$ between these two sets of resonances can be finely controlled by applying different level of stress on the FFP resonator [50]. For the DKS generation, a tunable external cavity diode laser followed by an erbium-doped fiber amplifier (EDFA) is coupled into the FFP resonator as the primary laser (Fig. 1(a)). When the primary laser is polarized along P1 with over 0.19 W power, SBL generation can be observed at the frequency downshifted from the primary laser by $f_{SBL}$ = 9.242 GHz. As we further increase the primary laser power to 4.3 W, the SBL will be efficient enough that it eventually overtakes the primary laser and becomes the dominant comb generating secondary pump in the resonator (inset of Fig. 1(c)). At this time, the intra-cavity power of the SBL will be 3 dB higher than that of the primary laser. Careful analysis of SBL and cavity mode structure reveals that the first-order Stokes Brillouin gain spectrum overlaps with the P2 resonance on the red-detuned side when the offset frequency $\Delta f_{P1-P2}$ is set to be 224 MHz. Thus, we can locate the SBL at the red-detuned side of the P2 resonance while the primary laser is still at the blue-detuned side of the P1 resonance [50]. As elaborated later in Fig. 2, such novel two-step pumping arrangement results in a self-stabilized strategy for thermal management and robust DKS generation. Of note, the choice for pump mode of P1 and P2 is arbitrary as long as the Brillouin shift frequency matches with the offset frequency of the corresponding cross-polarized cavity resonances by stress tuning. Importantly, the cascaded SBL is strictly forbidden in this cross-polarization arrangement as only the first-order Stokes Brillouin gain spectrum can ever overlap with the cavity resonance [50]. This is in stark contrast to the previous literature [39] where the cascaded SBL interacts with the cavity Kerr nonlinearity to enhance the hyperparametric oscillation but interrupt the DKS generation [51, 52]. Our configuration suppresses these unnecessary nonlinear interactions and promotes the generation of low-noise DKS.

When we further tune the P1-polarized primary laser into the resonance from the blue-detuned side, a P2-polarized DKS frequency comb centered around the SBL is obtained (DKS comb in Fig. 1(c)). Its 3-dB and 30-dB bandwidths are about 20 nm and 115 nm, respectively. The power of generated DKS comb is about 98 μW (corresponding to 0.1 pJ pulse energy) with conversion efficiency of $2.3\times10^{-5}$ from the primary laser. Following the development of the DKS frequency comb, we also observe a 10-dB weaker P1-polarized comb (pump comb in Fig. 1(c)) that is generated through cross-phase modulation (XPM) of the DKS [53]. The DKS is self-stabilized and it stays over hours without any active control. The DKS state is deterministic in that it can be repeatedly and reliably generated following the same pump tuning protocol. To elucidate the efficacy of the two-pumping scheme, we record the cavity transmission at both polarizations simultaneously while the primary laser frequency is scanned at a speed of 1.84 GHz/s across the P1 resonance from the blue-detuned side (Fig. 2(b)). The principle of the self-stabilized strategy for thermal management and robust DKS generation are illustrated in Fig. 2(c). When the primary laser is first tuned closer to the P1 resonance, the intracavity power gradually increases to above the cross-polarized SBL threshold for the initial SBL growth around the peak of the P2 resonance (stage I). When the primary laser is further tuned into the P1 resonance, the SBL secondary pump is pushed across the P2 resonance to the red-detuned side following the shift of the Brillouin gain peak (stage II). Of note, the SBL secondary pump experiences a much slower detuning speed with a slow-down factor of 15 compared to that of the primary laser, considering 5.6 MHz cavity resonance linewidth and 79 MHz Brillouin gain bandwidth [39, 54]. At the end of stage II, discrete transmission steps characteristic of DKS generation are observed. In general, such intracavity power drop leads to cavity cooling and consequently resonance blueshift that induces thermal instability [37, 38, 55]. In contrast, here the dynamics is very different as the primary laser is still on the blue-detuned side of the resonance. The resonance blueshift now leads to an increased cavity loading from the primary laser and it compensates for the intracavity power drop from the SBL secondary pump alone. Thus, thermal equilibrium is still maintained at the DKS state. Such self-stabilized strategy mitigates the thermal instability and facilitates robust DKS generation. At stage III, the primary laser continues to get closer to the P1 cavity resonance while the SBL secondary pump is pushed further away from the P2 cavity resonance, resulting in the monotonic increase and decrease of the P1 and P2 transmission signals respectively. Finally, primary laser is also scanned to the red-detuned side of the P1 resonance, resulting in a trivial thermal instability that quickly pushes the system off the resonance (stage IV).

Figure 3(a) plots the radio frequency (RF) spectra of the beat notes from the mixed polarization output (top panel) and the P2-polarized DKS (bottom panel). Clean RF beat notes with signal-to-noise ratio (SNR) greater than 45 dB at 1-MHz resolution bandwidth (RBW) are observed at harmonic frequencies. The frequency separation $\Delta f$ between each harmonic beat note and its nearest side peaks is measured to be 224 MHz, in a good agreement with the offset frequency $\Delta f_{P1\text{-}P2}$ between orthogonal modes in the FFP resonator. The RF measurement further shows that the two orthogonally polarized combs are synchronized with the same repetition frequency and the same narrow linewidth [50] within the instrument limit, which can be explained by the XPM coupling in between. Importantly, P2-polarized DKS frequency comb can be well isolated and selected by a combination of a half-wave plate (HWP) and a polarizing beam splitter (PBS) cube at the output. Figure 3(b) is the zoom-in view of its fundamental beat note, showing a resolution-limited linewidth of 10 Hz.

Our two-step pumping scheme not only mitigates the thermal instability challenge, but also provides a passive linewidth narrowing for the secondary pump through the cavity enhanced SBL process. It relaxes the laser linewidth requirement for the DKS to reach the quantum-limited phase noise and timing jitter. In this process, the SBL (or the secondary pump) has a much higher spectral purity than its pump (or the primary laser) with a linewidth narrowing factor of $(1 + \gamma_S/\Gamma)^2$ where $\Gamma$ and $\gamma_S$ are the cavity linewidth and the Brillouin gain bandwidth respectively [39]. In our FFP resonator where $\Gamma = 5.6$ MHz and $\gamma_S = 79$ MHz, a narrowing factor of 228 is expected such that the secondary pump linewidth can be reduced to 22 Hz from the 5-kHz primary laser. Experimentally, we build an SDSHI to characterize such significant linewidth narrowing. SDSHI uses short delay fibers and curve fitting of the measured coherent envelope to suppress the $1/f$ Gaussian noise and obtain a more accurate estimate of the intrinsic laser Lorentzian linewidth [44, 50]. Figure 3(c) plots the SDSHI experimentally measured results overlaid with theoretically calculated results of the primary laser (top panel) and the DKS frequency comb (bottom panel). A good agreement between the measurement and the calculation is obtained, and it confirms the linewidth narrowing capability of our two-step pumping scheme. Consequently, our DKS frequency comb has an unprecedented 22-Hz intrinsic linewidth that facilitates the achievement of quantum-limited phase noise and timing jitter into the attosecond realm.

The phase noise and timing jitter are analyzed with the ARMI timing jitter measurement method [50], which provides attosecond timing jitter precision with a -194 dBc/Hz measurement noise floor at 1-GHz carrier (equivalent to $2 \times 10^{-3}$ as$^2$/Hz)

[45]. For our FFP resonator, the quantum-limited phase noise should have a characteristic 20 dB/decade roll-off and approach a white noise floor of -182 dBc/Hz with a corner offset frequency of 1 MHz [33]. Such low phase noise floor cannot be accurately assessed by direct photodetection methods that are typically shot noise-limited at -160 dBc/Hz level [45, 56, 57]. Figure 3(d) plots the experimentally measured single sideband phase noise (SSB PN) spectrum overlaid with the theoretically calculated quantum limit. Excellent agreement between the measurement and the theory is obtained for offset frequencies above 10 kHz, except for a few high-frequency measurement artefact spikes resulting from the 45-m-long fiber link in the ARMI setup [45]. The integrated timing jitter is 2.5 fs for averaging times up to 100 μs, corresponding to less than half of the optical cycle at the DKS center wavelength. It further breaks into the attosecond timing jitter realm of 995 attosecond for averaging times up to 10 μs (Fig. 3(e)). For offset frequencies below 10 kHz, relative intensity noise (RIN) from the EDFA induced excessive noise through self-steepening dominates the SSB PN spectrum. It rapidly increases the integrated timing jitter to above the single-optical-cycle level for averaging times up to 1 ms. However, these low frequency noise components can be well suppressed by either passively isolating the DKS setup from the environment or actively controlling the DKS power through standard electronics. Here we choose 10 MHz Fourier frequency as the upper limit for the timing integration as the white noise floor above 10 MHz can also be significantly reduced by orders of magnitude through applying a narrow-band RF filter to replace the shot noise with thermal noise [58]. The measurement confirms the demonstrated DKS frequency comb, with the novel FFP resonator platform and two-step pumping scheme, achieves an unprecedentedly low timing jitter in comparison to other on-chip combs and complements the traditional mode-locked laser frequency comb as a compact photonic flywheel.

Finally, the DKS pulse structure is characterized using a second-harmonic generation frequency-resolved optical gating (SHG-FROG) system. A stable bright single-soliton pulse train with a repetition period of 1.06 ns is observed on the oscilloscope [50]. Before the pulse is directed to the SHG-FROG system, it passes through a free-space grating filter and then coupled into fiber with a coupling efficiency of 24 %, to have the primary laser and secondary pump blocked. Then a dispersion compensated EDFA is used to amplify the pulse energy to 740 pJ with spectrum shown in the inset of Fig. 4(b). The pulse shape and temporal phase are retrieved using an iterative genetic algorithm [50, 59] as plotted in Fig.4(a), showing a nearly transform-limited pulse duration of 220 fs with a FROG error of 0.6%. The pulse has enough peak power for high-quality supercontinuum generation in a 6-m long dispersion-flattened HNLF (HNDS1626BA, SUMITOMO ELECTRIC) with dispersion of 5.8 ps/nm/km and dispersion slope of 0.026 ps/nm$^2$/km at 1550 nm. The HNLF has a high nonlinear coefficient of 24 W$^{-1}$km$^{-1}$, and its zero-dispersion wavelength is extended to 1395 nm, facilitating the generation of Cherenkov radiation at shorter wavelengths around 1 μm. Overall, the supercontinuum spans over an octave (Fig. 4(b)) where 292-pJ pulse energy is concentrated within the 2000-2300 nm range to generate f-2f beating signal with enough signal-to-noise ratio. This signal can be fed back to the pump laser frequency through its piezo stage or the diode current, for future SHG frequency comb self-referencing [3].

In summary, we demonstrate for the first time a compact DKS photonic flywheel. The choice of FFP resonator platform enables quantum-limited sub-femtosecond timing jitter at gigahertz repetition rate, and the innovation of two-step pumping scheme is instrumental in mitigating the DKS thermal instability and enforcing the DKS generation at the quantum limit. Free-running, our DKS photonic flywheel achieves an intrinsic comb linewidth of 22 Hz and enters into the attosecond timing jitter realm of 995 attosecond for averaging times up to 10 μs. It has the benefit of DKS for considerably larger comb spacings in non-conventional spectral ranges but can achieve a performance that was only attainable previously in traditional mode-locked lasers. The spectrum centered at 1557 nm has a 3-dB bandwidth of 20 nm, and it can be further broadened via supercontinuum generation in dispersion-flattened HNLF to cover more than an octave. It founds the basis for frequency comb self-referencing toward even lower time jitter at all time scales. Broader comb spectra as well as shorter pulse duration can be expected for further fiber dispersion engineering. And the fabrication methods can be improved to achieve higher finesses and lower the pump power requirement. Finally, the intrinsic compatibility between our FFP resonator platform and the existing fiber laser technology will facilitate hermetic all-in-fiber packaging to achieve the long-sought-after goal of field-deployable precision metrology device in both spectral and time domains.

### Acknowledgements

We thank Prof. Vahala, K. J., Prof. Diddams, S. A., Dr. Coddington, I. R., and Prof. Gaeta, A. L. for discussions. K. J., X. W., H. L., X. N., J. G., Z. X., and S. Z. acknowledge the support by the National Key R&D Program of China (2019YFA0705000, 2017YFA0303700), Key R&D Program of Guangdong Province (2018B030329001), Leading-edge technology Program of Jiangsu Natural Science Foundation (BK20192001), National Natural Science Foundation of China (51890861, 11690031, 11621091, 11627810, 11674169, 91950206). K. J., X. W., D. K., J. W., E. T., H. L., and S.-W. H. acknowledge the support from the University of Colorado Boulder. K. J., X. W., and H. L. acknowledge the support by China Scholarship Council (CSC). J. K. and D. K. acknowledge the support by Institute of Information & Communications Technology Planning & Evaluation (IITP) of Korea (2019-0-01349) and National Research Foundation of Korea (2018R1A2B3001793).

* E-mail: xiezhenda@nju.edu.cn; ShuWei.Huang@colorado.edu

† These authors contributed equally to this work.

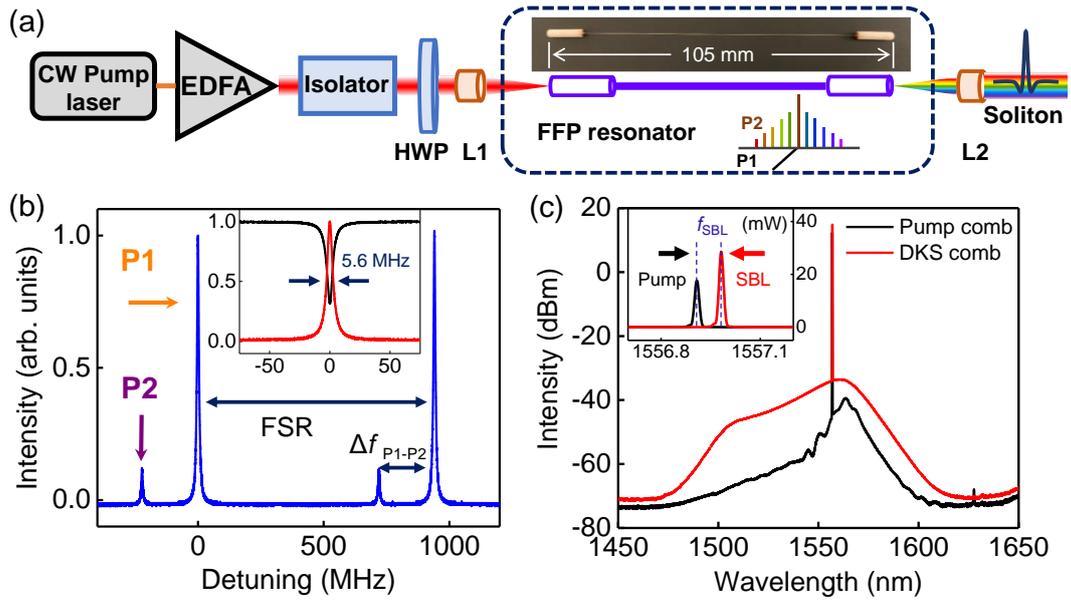

FIG. 1. (a) Experimental setup for self-stabilized DKS generation. L1 and L2: aspheric lens. Inset: picture of the FFP resonator. (b) Cold-cavity transmission spectrum around 1557 nm. Inset: Zoom-in view of the transmission and reflection signals across the P1 resonance. (c) When the primary laser is polarized along P1, efficient SBL generation (inset) followed by DKS frequency comb generation, both polarized along P2, is observed and measured.

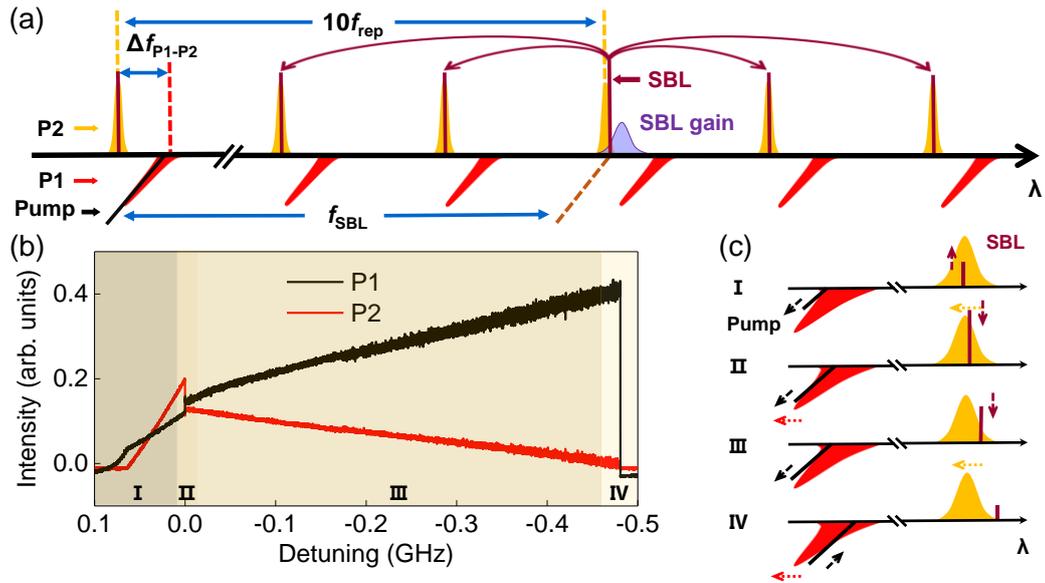

FIG. 2. (a) Schematic of the self-stabilized DKS generation. When the DKS is generated, the P1-polarized primary laser remains on the blue-detuned side while the P2-polarized SBL secondary pump is pushed to the red-detuned side of the cavity resonances. Such arrangement mitigates the thermal instability and facilitates robust DKS generation. (b) Cavity transmission at both polarizations during the primary laser scan from higher to lower frequencies over a resonance. (c) Principle of the self-stabilized strategy for thermal management and robust DKS generation. Stage I: SBL growth; Stage II: DKS generation; Stage III: chaotic oscillation; and Stage IV: trivial off-resonance state.

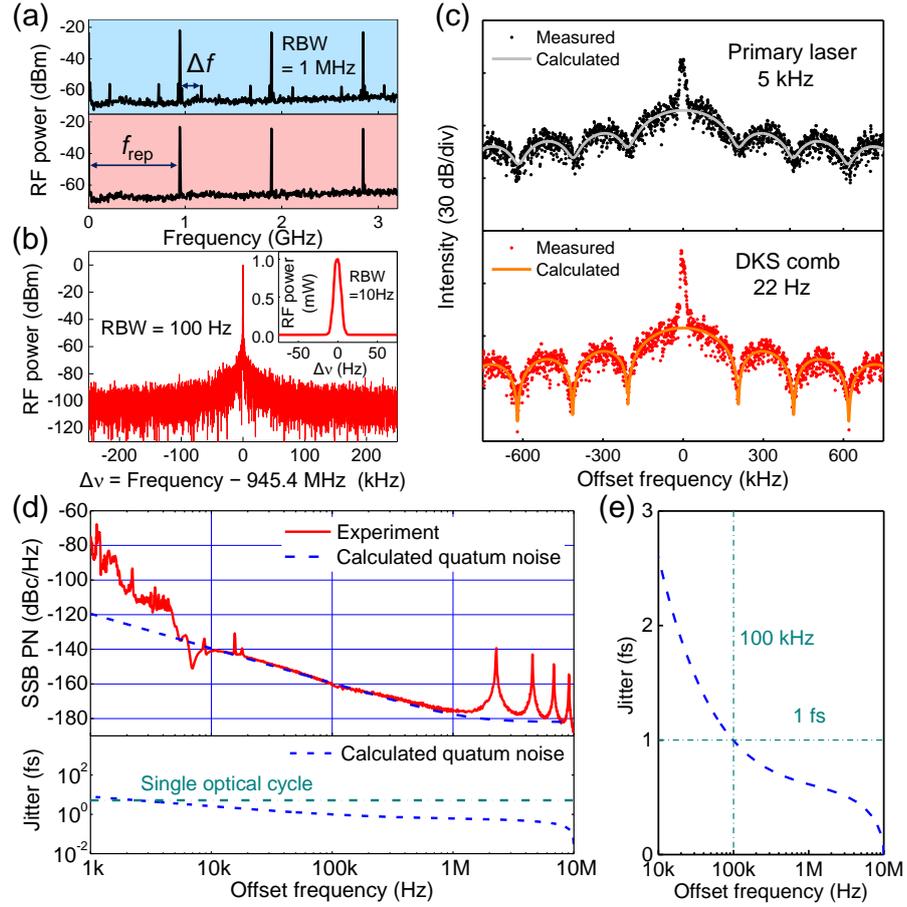

FIG. 3. (a) RF beat notes of the mixed polarization output (top panel) and the P2-polarized DKS frequency comb (bottom panel). (b) Fundamental beat note of the DKS frequency comb on logarithmic scale and linear scale (inset). (c) SDSHI results of the primary laser and the DKS frequency comb, showing a significant linewidth narrowing from 5 kHz to 22 Hz. (d) Top panel: single sideband phase noise of DKS frequency comb scaled at 1-FSR carrier (red solid curve), reaching the quantum noise limit (blue dashed curve) for offset frequencies above 10 kHz. Bottom panel: sub-optical-cycle integrated timing jitter. (e) Zoom-in view of the integrated timing jitter, showing a sub-femtosecond jitter for averaging times up to 10 μs.

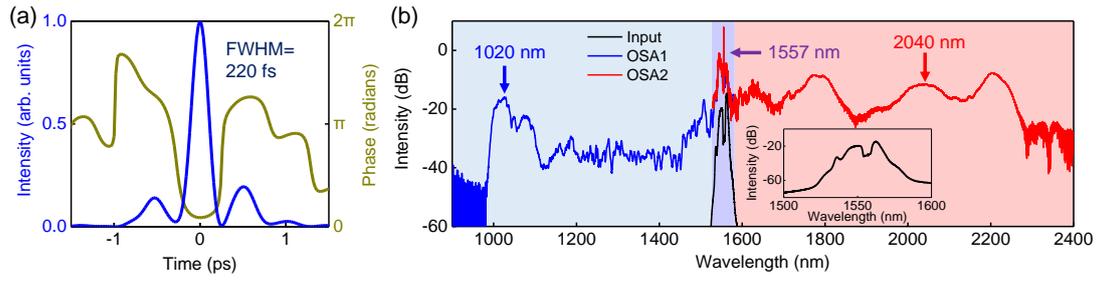

FIG. 4. (a) FROG-retrieved pulse shape and temporal phase. (b) Supercontinuum spanning more than an octave. Inset: spectrum of the amplified single-soliton pulse.

# Supplemental Material for
# "Photonic flywheel in a monolithic fiber resonator"

Kunpeng Jia[1,2]†, Xiaohan Wang[1,2]†, Dohyeon Kwon[3], Jiarong Wang[2], Eugene Tsao[2], Huaying Liu[1,2], Xin Ni[1], Jian Guo[1], Mufan Yang[1], Xiaoshun Jiang[1], Jungwon Kim[3], Shi-ning Zhu[1], Zhenda Xie[1]*, Shu-Wei Huang[2]*

[1] *National Laboratory of Solid State Microstructures, School of Electronic Science and Engineering, College of Engineering and Applied Sciences and School of Physics, Nanjing University, Nanjing 210093, China.*

[2] *Department of Electrical, Computer and Energy Engineering, University of Colorado Boulder, Boulder, CO 80309, United States.*

[3] *Department of Mechanical Engineering, Korea Advanced Institute of Science and Technology, Daejeon 34141, Republic of Korea.*

†These authors contributed equally to this work.
*Correspondence to: xiezhenda@nju.edu.cn; ShuWei.Huang@colorado.edu

## SBL generation and characterization
### 1. Stress control for SBL generation

In a fiber Fabry-Perot (FFP) resonator, the birefringence between the P1 and P2 resonances is caused by the fiber stress, so that their offset frequency can be tuned by varying the stress on the resonator [1]. This scheme is important for achieving cavity enhanced stimulated Brillouin lasing (SBL) in cross polarizations, which plays an essential role in our two-step pumped dissipative Kerr soliton (DKS) generation. As shown in Fig. S1(a), SBL can only be cavity enhanced and thus efficiently generated when a P2 resonance overlaps with the stimulated Brillouin gain spectrum, which can be achieved by stress tuning in non-polarization-maintaining fiber as we use in this work. This scheme can be further implemented to polarization-maintaining fibers whose frequency detuning between two polarization modes is set appropriately to meet the above requirement [2]. This may further improve the environmental stability of the system.

In the setup, we specially designed a metal mount for this stress tuning. This mount is composed of two metal plates to sandwich the FFP resonator, as shown in Fig. S1(b). These metal plates are finely machined with a groove to fit and grab the FFP resonator in-between. A set of screws are used to adjust the tension of spring pairs which finely tunes the stress and thus the P2 to P1 resonance shift. This stress tuning is also important to achieve effective red-detuning for the SBL to the P2 resonance in a thermally stable state. As shown in Fig. S1(b), it is achieved by putting the P2 resonance on the blue side of the SBL gain.

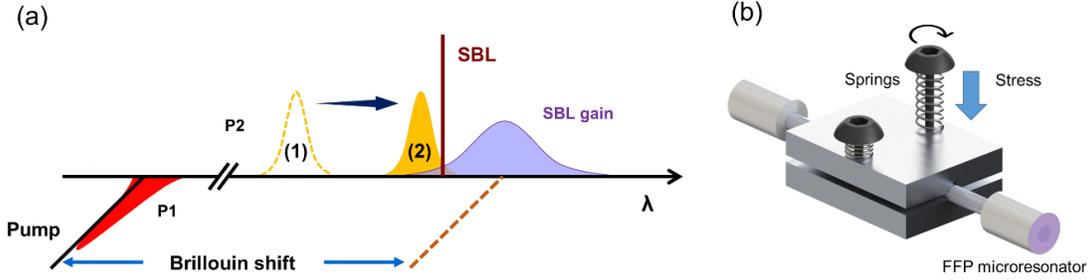

FIG. S1. Stress tuning for SBL generation. (a) Schematic of resonance tuning for SBL generation. The stress on fiber changes the modal birefringence of P1 and P2 modes, leading to a change of offset frequency between P1 and P2 resonances. The solid P2 resonance marked in orange shows an overlap with the Brillouin gain spectrum on its blue side, which is optimized for the effectively red-detuned SBL generation, as required for our two-step pumped DKS generation. (b) FFP resonator mount. Two pairs of screw/spring sets are used to add stress on the resonator. The fine threads of the screws result in a fine tuning of the stress. In addition, the base plate temperature is actively controlled with a 10 mK stability.

## 2. SBL frequency shift measurement

In cavity, the frequency shift of SBL from pump is decided by the cavity resonance and SBL gain spectrum together. Here with DKS generated, we use a band pass filter to filter out the pump and SBL and down mix their beat note signals with a 10 GHz local oscillator for baseband detection in an electrical spectrum analyzer (ESA).

An RF signal at 758 MHz is measured, which corresponds to a 9.242 GHz frequency shift between the pump and the SBL peak (9.242 GHz = 10 GHz − 758 MHz). Considering the offset frequency between two sets of cavity modes ($\Delta f_{P1-P2}$ = 224 MHz), the SBL frequency shift is 12 MHz larger than $10f_{rep} - \Delta f_{P1-P2}$ (9.230 GHz), which is contributed to the small blue-detuning of the primary laser and the red detuning of the SBL secondary pump as shown in Fig. S2.

Of note, the frequency shift between SBL gain from pump is fixed in the FFP resonator, so if the primary laser polarization is aligned along P2, no SBL can be generated as the Brillouin gain spectrum cannot overlap with any resonance considering its narrow bandwidth (~100 MHz).

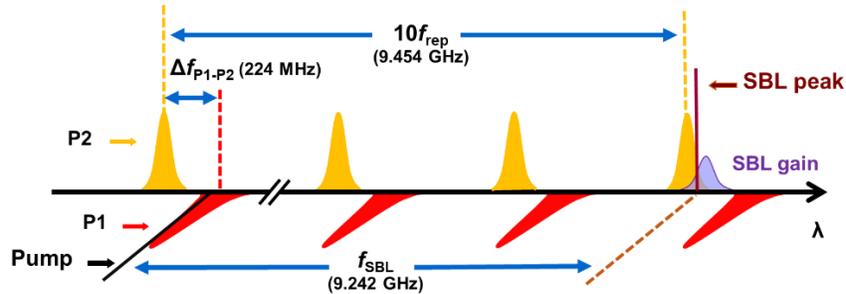

FIG. S2. Schematic diagram of SBL frequency shift.

## 3. Evolution of SBL peak

To study the evolution of SBL in our DKS generation process, the spectra of SBL peak are recorded when we slightly tune the pump into the P1 resonance from its blue side. As shown in Fig. S3, SBL is excited once the intra-cavity pump power exceeds SBL threshold, and it gets stronger with the pump approaching P1 resonance. SBL eventually exceeds the pump in intensity and dominants comb generation. It's worth noting that there is no observation of cascaded SBL and mixing with pump, as these processes cannot be overlapped with cavity resonances and enhanced by the cavity.

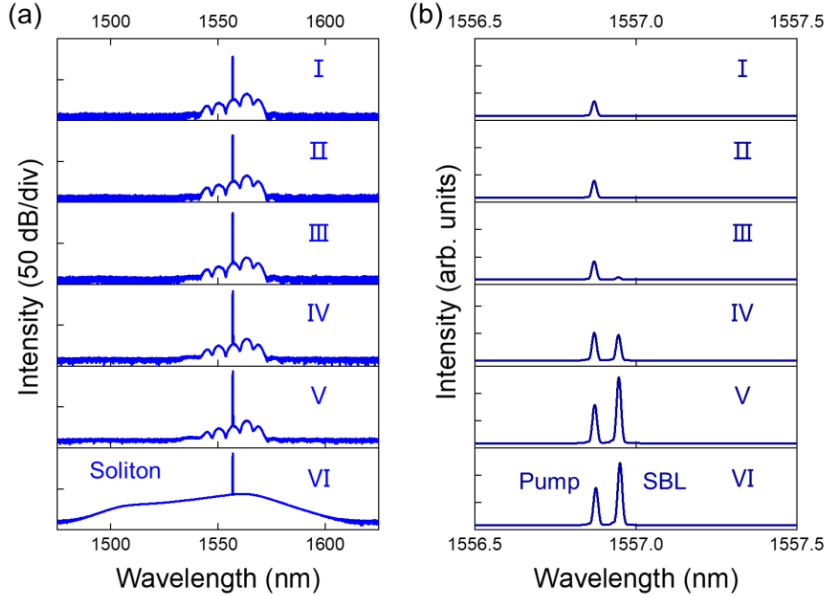

FIG. S3. Spectra of SBL peak evolution. Pump light is tuned into P1 resonance with decreasing blue detuning from I to VI. SBL is excited at state III and gradually grows with pump going deeper into the cavity resonance. SBL starts to overtake the pump at state V and dominate the comb generation at state VI. (a) Full span spectra. (b) Zoom-in spectra around pump wavelength.

Dispersion measurement of the HNLF

Our FFP resonator is made of highly nonlinear fiber (HNLF) and the dispersion of such HNLF is measured by the spectral interferometric method [3]. Here a broad-spectrum laser (Menlo System) is used. As shown in Fig. S4, the light is directed in a Michelson interferometer, where the upper arm contains 0.5 m-long HNLF in a back-coupled double-pass scheme, and the lower reference arm is in free-space only with a movable reflective mirror. It is used to precisely adjust the optical path length difference between these two arms. Reflected light beams in the two arms coincide together through the 50/50 beam splitter and are collected into a single mode fiber and sent to an optical spectrum analyzer (OSA). The dispersion of the HNLF can be fitted from their spectral interference pattern as shown in Fig. S5(a). We apply a digital high-pass filter firstly to remove the low frequency modulation for easier fitting (Fig. S5(a)).

The interference pattern can be fitted with
$$I(\omega) \propto 1 + \cos[\varphi(\omega)] \tag{S1}$$
here $\varphi(\omega)$ is the phase difference of the two arms
$$\begin{aligned}\varphi(\omega) &= 2(\beta_{arm1}(\omega)L_{arm1} - \beta_{arm2}(\omega)L_{arm2}) \\ &= 2[\beta_{air}(\omega)L_{air1} + \beta_{HNLF}(\omega)L_{HNLF} - \beta_{air}(\omega)L_{air2}] \\ &= 2[\beta_{HNLF}(\omega)L_{HNLF} - \beta_{air}(\omega)\Delta L_{air}]\end{aligned} \tag{S2}$$

$\beta_{HNLF}(\omega)$ and $\beta_{air}(\omega)$ are the propagation constants in HNLF and air respectively. By Taylor expansion at $\omega_0$ and ignoring the high order dispersion in the air, we can get

$$\varphi(\omega) = \varphi_0 + 2[(\beta_{HNLF1}(\omega_0)L_{HNLF} - \frac{\Delta L_{air}}{c})(\omega - \omega_0)]$$

$$+ 2 \times \frac{1}{2}\beta_{HNLF2}(\omega_0)L_{HNLF}(\omega - \omega_0)^2$$

$$+ 2 \times \frac{1}{6}\beta_{HNLF3}(\omega_0)L_{HNLF}(\omega - \omega_0)^3 \quad (S3)$$

$\beta_{HNLF1}$, $\beta_{HNLF2}$, $\beta_{HNLF3}$ are the first, second and third order dispersion. Here the $\beta_{HNLF2}$ at 1,557 nm of the HNLF is fitted to be about -3 fs$^2$/mm, showing a small anomalous group velocity dispersion (GVD).

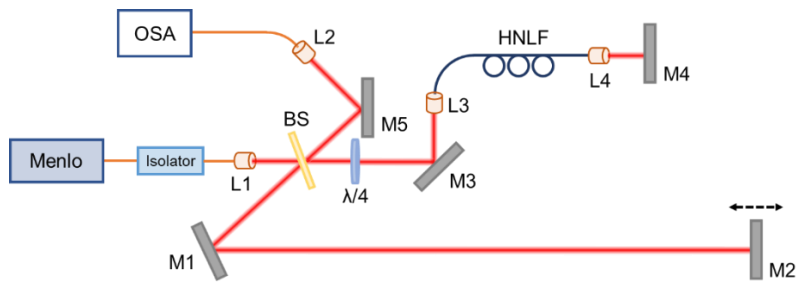

FIG. S4. Experimental setup for dispersion measurement. L1, L2, L3 and L4: aspheric lenses; BS: 50/50 beam splitter; λ/4: quarter-wave plate, used in one of the arms to ensure the maximum interference intensity. M1, M2, M3, M4 and M5: reflective mirrors, here M2 is mounted on a translation stage which can travel along the light path. OSA: optical spectrum analyzer.

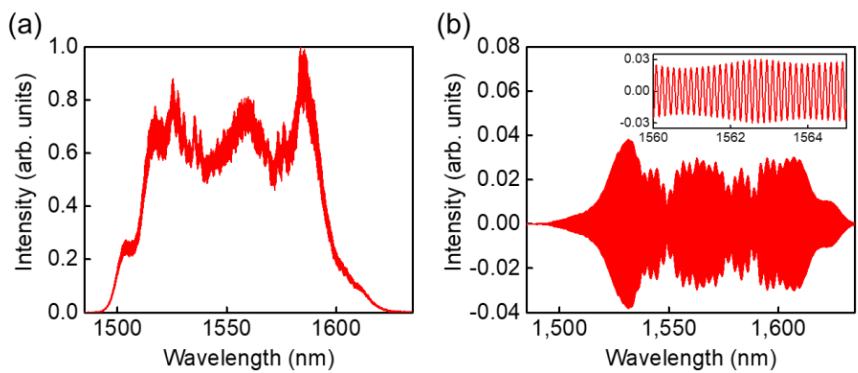

FIG. S5. Interference spectrum. (a) Measured interference spectrum. (b) Interference spectrum after FFT filter. The inset shows the zoom-in interference pattern.

DKS frequency comb linewidth measurement

Conventional decoherence laser linewidth measurement methods like delayed self-heterodyne/homodyne interferometry require a fiber delay length that is comparable to the laser coherence length. For our DKS comb, a comb linewidth of 22 Hz is expected and thus delayed fiber length of ~10$^4$ km is needed for a decoherence operation, which is unpractical in experiment and may induce unfavorable linewidth broadening in the long delay fiber. On the other hand, strong coherent envelope can also be used to characterize the laser's intrinsic linewidth through self-coherent detection [4]. The laser linewidth can be fitted from the self-coherent envelope of laser, and relatively short delayed fiber is needed (~1 km).

Here we characterize DKS comb linewidth by fitting the strong coherent envelope through self-coherent detection. The setup is shown in Fig. S6, consisting of 1 km delayed fiber. After amplified in an EDFA, the DKS comb is divided into two arms by a 1×2 fiber coupler. The comb in one arm is frequency-shifted in a 200 MHz acoustic optical modulator, while the comb in another arm is delayed by 1 km single-mode fiber. Polarization controllers (PC) are used to ensure maximum interference intensity. These two arms are recombined by another 1×2 fiber coupler. Their beating signal is then detected by a PD and measured using an ESA.

The detected beating signal can be characterized by the following equations

$$S(f,\Delta f) = S_1 S_2 + S_3 \tag{S4}$$

here

$$S_1 = \frac{P_0^2}{4\pi} \frac{\Delta f}{\Delta f^2 + (f - f_1)^2} \tag{S5}$$

$$S_2 = 1 - e^{-2\pi\Delta f \tau_d} \left[ \cos[2\pi(f \pm f_1)\tau_d] + \Delta f \frac{\sin[2\pi(f \pm f_1)\tau_d]}{f \pm f_1} \right] \tag{S6}$$

$$S_3 = \frac{\pi P_0^2}{2} e^{-2\pi\Delta f \tau_d} \delta(f \pm f_1) \tag{S7}$$

here $P_0$ is the detected optical power; $\Delta f$ is the Lorentzian linewidth; $f$ is the measurement frequency; $f_1$ is the AOM frequency shift (here we use 200 MHz); $\tau_d = L/c$, is the time delay of one arm with respect to another, where $L$ is the optical length of the delayed fiber, $c$ is the speed of light. For $S_3$, when $f \neq f_1$, $\delta(f \pm f_1) = 0$, $S_3 = 0$ and when $f = f_1$, $S_3$ = infinite. The electric spectrum $S$ will be unstable at $f = f_1$, so we use $S(f, \Delta f) = S_1 S_2$ for the coherent envelope fitting. The measured beating signal and calculation result are shown in Fig. 3(c) of the main text.

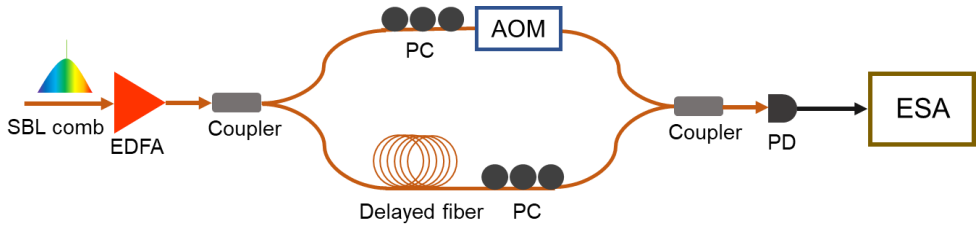

FIG. S6. Linewidth measurement setup. EDFA: erbium-doped fiber amplifier; PC: polarization controller; AOM: acoustic optical modulator; PD: photo detector; ESA: electric spectrum analyzer. Here the length of the delayed fiber is 1 km.

Time-domain characterization of single soliton

Fig. S7(a) shows the bright single-soliton pulse train with a repetition period of 1.06 ns measured by an oscilloscope. We characterize the temporal feature of single soliton by a second-harmonic generation frequency-resolved optical gating (SHG-FROG) system as described in the main text. The spectrum entering the SHG-FROG system is shown in Fig. S7(b) and the measured SHG-FROG spectrogram and autocorrelation result are shown in Fig. S7(c).

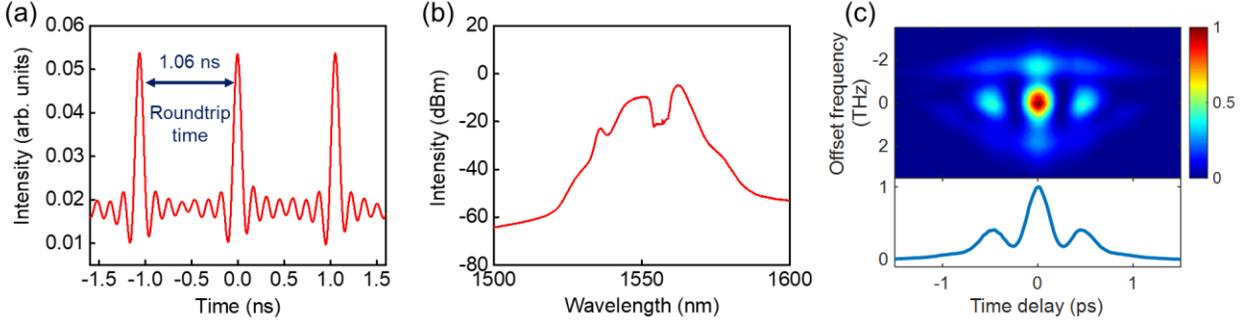

FIG. S7. (a) Bright DKS pulse train with a fundamental repetition period of 1.06 ns. (b) The pulse spectrum entering the SHG-FROG system after amplified in a dispersion compensated flat-gain EDFA. The primary laser and secondary pump have been blocked by a free-space grating filter. (c) Measured SHG-FROG spectrogram (top panel) and delay marginal (bottom panel).

## Other comb states
### 1. P1-polarized "pump comb" state

In our two-step pumping scheme, a weaker comb polarized along P1 and centered around the primary laser is also generated via cross-phase modulation (XPM) from the P2 polarized DKS comb. We name it pump comb and its spectrum is shown in Fig. 1(c) of the main text. This pump comb can be well separated from the DKS comb with a half-wave plate and polarizing beam splitter and its RF beat note signal is measured as shown in Fig. S8. The RF spectrum also shows a clean peak and relativity high signal-to-noise ratio (SNR) of 80 dB, indicating a different state from chaotic comb state.

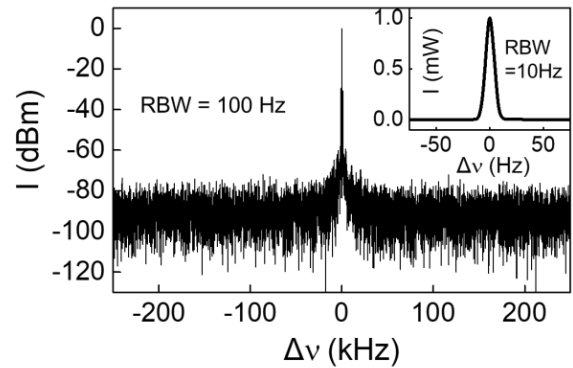

FIG. S8. RF beat note measurement of pump comb at repetition rate (945.4 MHz). Here $\Delta \nu$ = Frequency − 945.4 MHz. The vertical axis (I) represents intensity.

### 2. Multiple-soliton states

Different soliton states can be observed and self-stabilized from our FFP resonator, with passive stability as good as single soliton state. Here we present some spectra of multiple solitons in Fig. S9. The intensities of these multiple solitons are higher than that of single soliton and their spectra have different interference patterns induced by different number of solitons.

We characterize the temporal feature of a multiple soliton shown in Fig. S9(b) by a second-harmonic generation frequency-resolved optical gating (SHG-FROG) system as described in the main text. The spectrum entering the SHG-FROG system is shown in Fig. S9(g) and the measured SHG-FROG spectrogram and autocorrelation result are shown in Fig. S9(h).

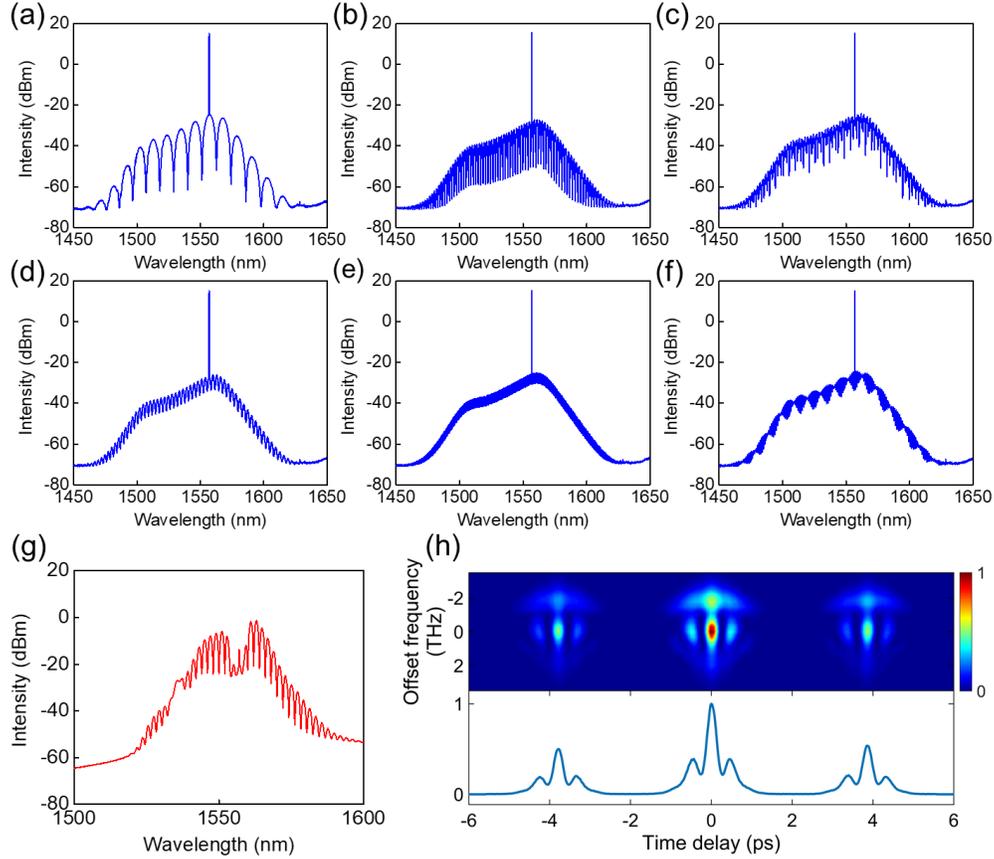

FIG. S9. (a-f) Experimental spectra of multiple solitons. (g) Multiple soliton spectrum for FROG measurement. (h) Measured FROG trace (upper) and autocorrelation (lower).

### 3. Chaotic comb states

We also measure the spectra and corresponding RF signals of chaotic comb states with pumping at P1 (Fig. S10) and P2 (Fig. S11), respectively, with different blue detuning of pump frequency from corresponding resonances. The P1-pumped spectra show different profile from P2-pumped ones, with lower RF signal noise floor and higher SNR, which may be attributed to the SBL noise reduction mechanism.

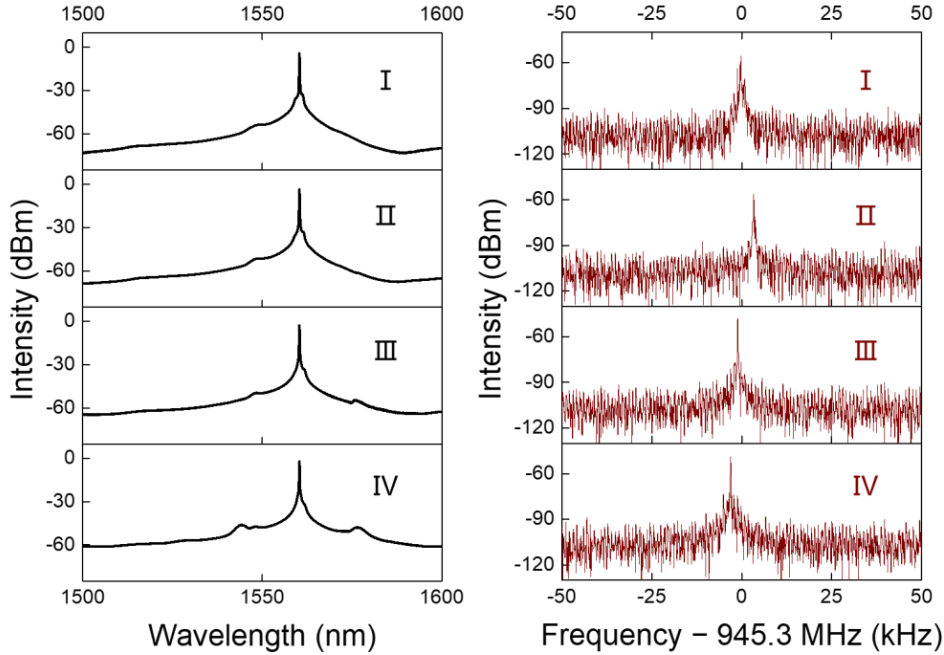

FIG. S10. Spectra and RF signals of P1-pumped chaotic comb states. Figures from I to IV correspond to decreasing blue detuning of pump frequency from resonance. The resolution bandwidth (RBW) in RF signals measurement is 100 Hz.

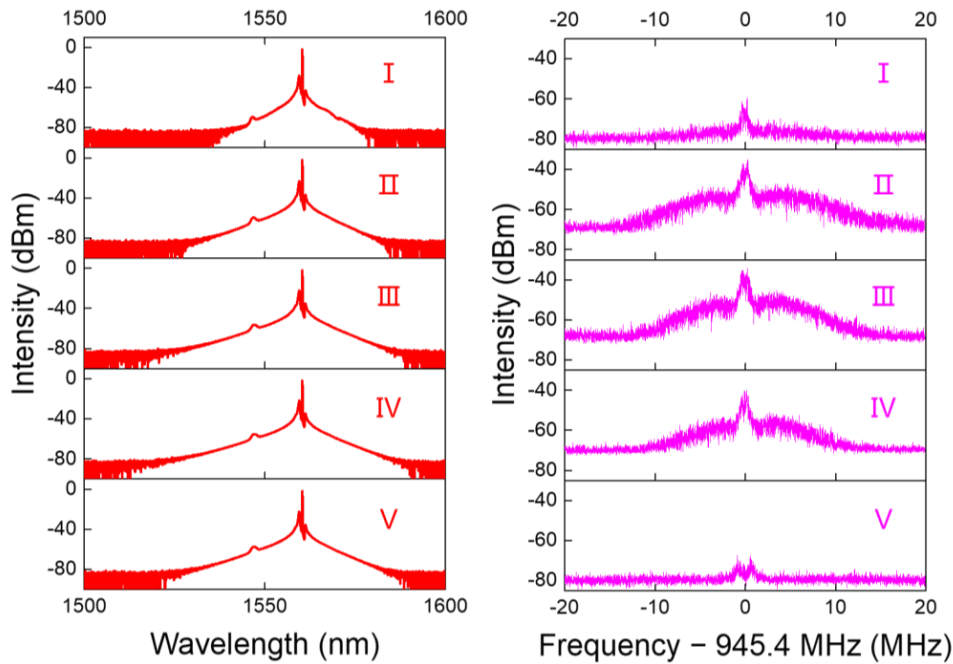

FIG. S11. Spectra and RF signals of P2-pumped chaotic comb states. Figures from I to V correspond to decreasing blue detuning of pump frequency from resonance. The resolution bandwidth (RBW) in RF signals measurement is 40 kHz.

Effect of frequency tuning speed in the thermal nonlinear dynamics

For the study of thermal nonlinear dynamics, we record the transmission intensities at different frequency decreasing speed of P1-polarized pump. Both P1 and P2 intensities are recorded for comparison, with pump power fixed at 4.3 W, as shown in Fig. S12. As described in the main text, when SBL reaches the red side of the P2 resonance, sudden decrease and increase on the intra-cavity power of SBL and pump, respectively, can be observed due to the thermal effect of cavity. With higher scan speed ($> 43.2$ GHz/s), the pump is tuned to deeper resonance thus more intra-cavity pump power right after the blue shift of resonance due to the SBL intensity drop, presented as a sharper peak followed by chaotic comb state in the P1 transmission. When we further increase the scan speed to over 835 GHz/s, the chaotic comb state is no longer accessible, as the thermal responds of cavity cannot follow the change of pump frequency. In this condition, the pump cross the resonance rapidly when the blue-shifted resonance encounters with the pump.

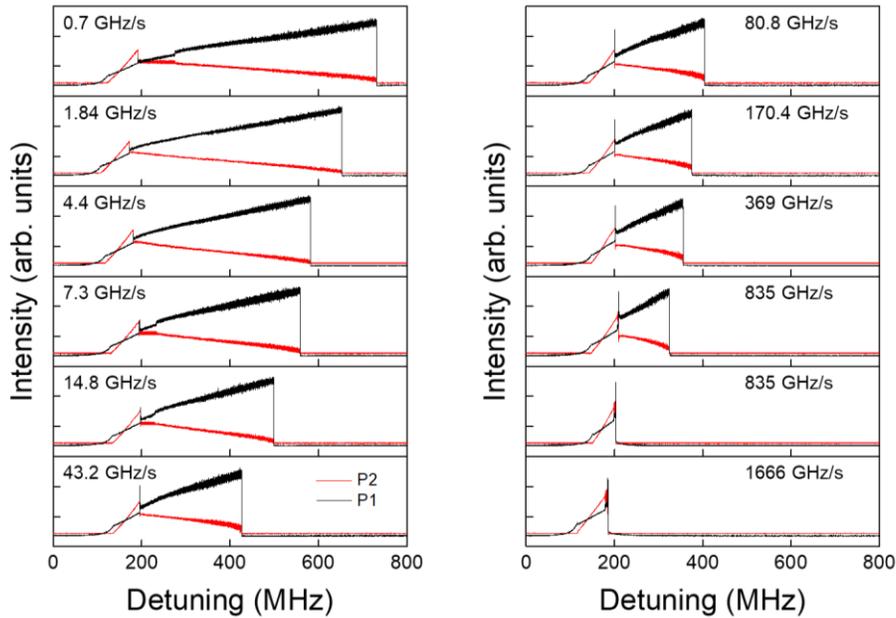

FIG. S12. P1-pumped cavity transmission with different sweeping speed.

Optimization of the FFP resonator

The FFP resonator used in our experiment is fabricated by mechanical polishing with optical dielectric Bragg mirror coating on both fiber ends. The intra-cavity loss of FFP resonator mainly originates from the diffraction at the end faces, as the propagation loss in fiber is negligible. The transmission/reflectivity coefficient of FFP resonator can be finely controlled by coating design and it is directly linked to the thickness of dielectric coating. An ideal reflector is supposed to have a thickness of zero, however, limited by the finite refractive indices contrast of coating materials, coating thickness of several micrometers is necessary to ensure a high reflectivity ($> 99.8\%$) thus a high-Q FFP resonator. Thus, diffraction loss in the coating is not negligible, and it dominates the intracavity loss of our FFP resonator.

To reduce the diffraction loss in the coating, a double-concave-cavity design may be applied in the future. The fiber end can be polished with a curved surface for better confinements of the reflected light. Here in the simulation, we use transmission matrix and overlap integral methods [5] to evaluate the diffraction loss at the fiber ends. As a comparison, the diffraction loss is approximately 0.26 % with a flat fiber end, while this value can be reduced to be close to zero with a surface radius of 0.23 mm. In this case, FFP resonator with higher Q can be expected.

### Frequency comb generation in a 10 GHz FFP resonator

From the fabrication point of view, the length of fiber resonator can be easily tailored to cover frequency range from 1 GHz to 10 GHz, which is of special interest for microwave photonics. Here we also demonstrate the frequency comb generation in an FFP resonator with a free spectral range (FSR) of 10 GHz. The HNLF is 10 mm in length and mounted in a ceramic ferrule. Both end faces are finely polished and coated with dielectric mirror with high reflection (R > 99.8 %) around 1,550 nm. The linewidth of resonance is measured to be 12.2 MHz as shown in Fig. S13(b), which corresponds to a Q factor of $1.6 \times 10^7$. The frequency comb is generated utilizing the filter-driven four-wave mixing method [6]. We can achieve frequency combs with comb spacing of 10 GHz and 40 GHz, corresponding to 1 FSR and 4 FSR (Figs. S13(c) and S13(d)), respectively.

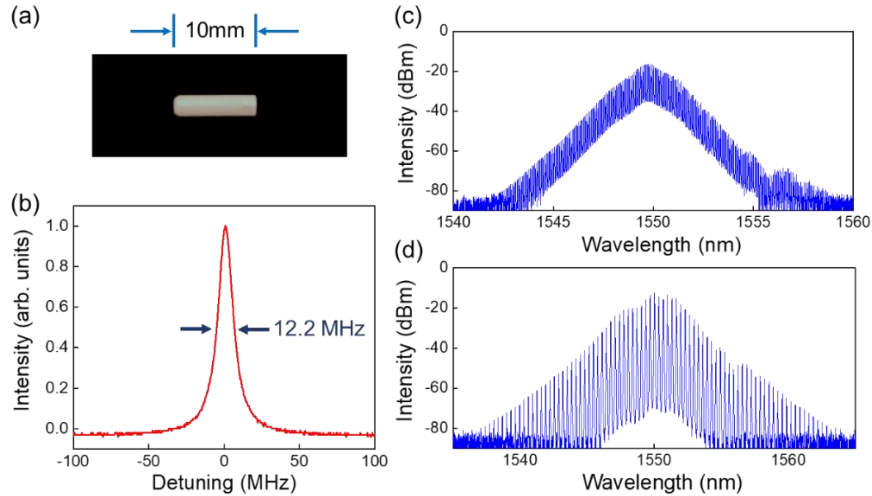

FIG. S13. 10 GHz FFP resonator and the comb generation. (a) Picture of a 10 mm-long FFP resonator. (b) The transmission signal when scanning a tunable laser across a resonance around 1550 nm. The linewidth is measured to be 12.2 MHz, which corresponds to a Q factor of $1.6 \times 10^7$. (c, d) Spectra with comb spacing of about 10 GHz and 40 GHz based on filter-driven four-wave mixing, respectively.